\documentclass[12pt]{article}
\usepackage{graphicx}
\usepackage{pazh}

\usepackage{lscape}
\usepackage{amssymb}

\tightenlines

\def\arcsec{$^{\prime\prime\,}$}
\def\arcmin{$^{\prime\,}$}

\def\a{$^{\mbox{\small a}}$}
\def\b{$^{\mbox{\small b}}$}
\def\c{$^{\mbox{\small c}}$}

\begin{document}

{\footnotesize { \it { Astronomy Letters, Vol. 33, No. 3, 2007, pp. 135-143. 
Translated from Pis'ma v Astronomicheskii Zhurnal, Vol. 33, No. 3, 
2007, pp. 186-195.
Original Russian Text Copyright \copyright\, 2007 by Karasev, 
Lutovinov, Grebenev.}}

\vspace{2mm}
\normalsize

\title{\bf Study of the Fast X-Ray Transient XTE J1901+014 Based on INTEGRAL, RXTE and ROSAT Data}

\author{\bf \hspace{-1.3cm}\copyright\,2007 . \ \
D.I.Karasev\affilmark{1}$^{\,*}$,
A.A.Lutovinov\affilmark{1},
S.A.Grebenev\affilmark{1}}
\affil{
$^1$ {\it Space Research Institute, Russian Academy of Sciences, Profsoyuznaya ul. 84/32, Moscow 117997, Russia} \\
}

\vspace{2mm}

\sloppy
\vspace{2mm}
\noindent

\sloppy

The source XTE J1901+014 discovered by the RXTE observatory during an
intense outburst of hard radiation and classified as a fast X-ray
transient is studied. The source's spectral characteristics in the
quiescent state have been investigated for the first time both in the
soft X-ray energy range (0.6-20 keV) based on ROSAT and RXTE data and
in the hard energy range ($>20$ keV) based on INTEGRAL data. A timing
analysis of the source's properties has revealed weak nonperiodic
bursts of activity on time scales of several tens of seconds and two
intense ($\sim$0.5-1 Crab) outbursts more than several hundred seconds
in duration. Certain assumptions about the nature of the object under
study are made.


\vfill

{$^{*}$ E-mail: dkarasev@hea.iki.rssi.ru}

\vfill

\section*{INTRODUCTION}

    XTE J1901+014 was discovered by the All-Sky
Monitor (ASM) of the RXTE space observatory during the outburst of April 6, 2002, when the photon flux from this source reached 0.9-1.2 Crab in
the energy range 1.5-12 keV. Because of technical
peculiarities of the ASM observations, it proved to
be possible to determine only the limiting outburst
duration interval:  $> 2$ min and $<3.15$ h. The J2000.0
coordinates of the source are R.A. = 19$^h$ 01$^m$ 46$^s$ and
DEC = +1\deg 24\arcmin 15\arcsec.7; the localization accuracy is  $\sim3$\arcmin
(Remillard and Smith 2002). No conclusion about
the nature of the source was reached, but it was
pointed out that the time profile of the outburst was
similar to the time profiles of the outbursts observed
from the microquasar V4641 Sgr. When analyzing
the archival ASM data, Remillard and Smith (2002)
pointed out the presence of a previous outburst from
the object under study occurred in July 1997 with a
flux of $0.4-0.5$ Crab in the energy range 1.5-12 keV
and a duration of $>6$ min and $<8$ h. No other equally
intense outbursts from the source have been detected.
  
  Using the ROSAT catalog, Remillard and Smith
(2002) established that 1RXS J190141.0+012618
localized by the PSPC instrument with the coordinates R.A. = 19$^h$ 01$^m$ 41$^s$ and DEC = +1\deg 26\arcmin 18\arcsec
is located within the ASM/RXTE error region of
the source XTE J1901+014. Analysis of the archival
ROSAT/HRI data showed that
1RXS J190141.0+012618 was also detected at a
statistically significant level during the observing
session of October 3, 1994. However, its position
was slightly displaced from the position initially
determined by ROSAT/PSPC (the ROSAT/HRI
position of the source is R.A. = 19$^h$ 01$^m$ 40$^s$.1 and DEC = +1 26\arcmin 30\arcsec ; the localization error is 10\arcsec ). It
was assumed that 1RXS J190141.0+012618 and,
after a refinement of the position using
ROSAT/HRI-1RXH J190140.1+012630, and
XTE J1901+014 are the same object, or the ROSAT
source is the nonflaring counterpart of the transient XTE J1901+014 (Wijnands 2002). However,
because of the relatively large uncertainty in the
position of XTEJ 1901+014, this assumption needed
additional testing.

    In this paper, we analyze a large number of observations of XTE J1901+014 at different times by
the INTEGRAL, RXTE, and ROSAT observatories,
refine the position of the source, and obtain and investigate the energy spectrum of the source in the quiescent state over a wide energy range, $0.6-100$ keV, and
its light curves for the first time. Preliminary results
of this study were published previously (Karasev et al. 2006).

\section*{OBSERVATIONS}
   In this paper, we use data from the ISGRI detector (Lebrun et al. 2003) of the IBIS gamma-ray
telescope onboard the INTEGRAL space observatory
(Winkler et al. 2003), the ASM monitor and the 
spectrometer onboard the RXTE space observatory
(Bradt et al. 1993), and the PSPC-C telescope on-board the ROSAT space observatory.

    The effective energy range of the ISGRI detector is from 15 to 200 keV, its angular resolution is
12 arcmin, and the localization accuracy of point
sources varies between $\sim$30\arcsec and several arcmin,
depending on their intensity. Other telescopes and
detectors of the INTEGRAL observatory (JEM-X,
PICsIT/IBIS, and SPI) failed to detect the source
at a statistically significant level. The operating energy ranges of the PCA/RXTE spectrometer and the
ASM/RXTE monitor are 3-20 keV and 1.3-12.2 keV, respectively; the sensitivity of the latter is $\sim$20 mCrab.
Because of the short exposure time, the source was
not detected by HEXTE/RXTE at a statistically significant level. The PSPC-C/ROSAT telescope has a
57\arcmin field of view and is capable of detecting photons
with energies 0.1-2.4 keV.

    The method of image reconstruction and spectral
analysis of the ISGRI data used in this paper was
described by Revnivtsev et al. (2004) and Lutovinov
et al. (2003). We reduced the observations with the
RXTE and ROSAT instruments using the standard
set of codes included in the HEASOFT 6.0 software package. The calibration data and the response
matrix for PSPC-C/ROSAT were taken from the
standard CALDB library. The background model for
 was chosen by taking into account the fact that
a weak source was studied.
    Table 1 lists the dates, observation numbers, and
orbits that were used in our analysis as well as
the ISGRI, PCA, and PSPC-C exposure times.
The publicly accessible ASM/RXTE data for the
source under study obtained from January 1996
through January 2006 were taken from the archive
at http://xte.mit.edu.

    We reduced the PSPC-C/ROSAT all-sky survey
data in accordance with the standard PSPC data
reduction technique for sources in the field of view of
the instrument slightly displaced from its center and
using recommendations from Belloni et al. (1994).

    When reducing the PCA/RXTE data, we took
into account the change in the source's detection
efficiency with its position in the field of view of the
instrument related to the collimator peculiarities (Jahoda et al. 2006). The deviation of the optical axis of
the spectrometer from the direction toward the source
varied between $\sim$1.2\arcmin and $\sim$7\arcmin during the observations;
this deviation was corrected by multiplying the collimator efficiency by an appropriate conversion factor.
It should be noted that the PCU4 detector module
of the PCA spectrometer was switched off during
the entire observation no. 30186; only three (PCU0,
PCU2, PCU3) of the five detector modules operated
during observation no. 70409. Although PCU0 has
had no propane veto layer, which greatly reduced the
background effect on the results obtained, since 2000,
we used its data in our analysis, because a separate
analysis of the behavior of this detector module revealed no significant anomalies that could distort the
results of observations. Note also that we used data
only from the two upper anode layers for each PCU.

      When the spectrum of the source under study
was reconstructed, it was necessary to take into account the influence of other sources that fell within the PCA field of view during certain observations and the Galactic ridge X-ray emission, because XTE J1901+014 (with Galactic coordinates:
$l= 35.38\deg$, $b=-1.62\deg$) falls into a region where its
influence is significant (Revnivtsev et al. 2006).
   
 In observation 30186-01-16-05S, apart from
XTE J1901+014, the X-ray pulsar XTE J1858+034
with a deviation of 59\arcmin.332 from the optical axis fell
within the PCA field of view. To properly reconstruct
the spectrum of the source under study from the combined spectrum of the sources based on observation
30186-01-16-05S, it was necessary to subtract the
spectrum of the pulsar XTE J1858+034. Assuming
that the shape of the latter remained unchanged,
we calculated it using observation 30137-01-01-15S, during which, according to the ASM data, the
pulsar's intensity was the same as that during the
observation of XTE J1901+014. We subtracted the
pulsar's spectrum by taking into account the PCA
collimator efficiency for a deviation of 59\arcmin.332 from the
optical axis (Jahoda et al. 2006); as a result, the flux
from it was multiplied by a factor of 0.03.
  
  To taken into account the influence of the Galactic
ridge X-ray emission, we used the fact that the time
of observation 30186-01-16-05S is divided into two
parts: slewing to the object and observation of the
object itself. During the slewing, the PCA axis moved
along the Galactic parallel b=$\sim-1.5\deg$, with the longitude changed from $\sim14\deg$ to $\sim35\deg$ (Fig. 1). Having 
chosen the times when no point sources fell within
the PCA field of view, we mark the level of persistent
radiation at this parallel that is the ridge X-ray emission. However, since the duration of one slewing is too
short to construct a statistically significant spectrum
of the ridge X-ray emission, an additional analysis of
the emission near the source as close to the Galactic
plane as possible, since its intensity is higher there,
was required. The validity of this approach follows
from the results of Revnivtsev et al. (2006), who point
out that the shape of the spectrum of the ridge X-
ray emission is constant at different Galactic latitudes
and longitudes. This is because this emission has the
same nature in any part of the galaxy. In addition, the
intensity of the ridge X-ray emission depends weakly
on longitude in the longitude range under consideration.
 
   To perform the required analysis of the ridge X-ray
emission, we can use data from slewings, pointings to
source-free regions, or pointings to transient sources
in the off state. We used the latter method, because
the best exposure was achieved in this case and, as
a result, the most significant spectrum of the ridge
X-ray emission was obtained. For our analysis, we
took observation no. 30141-05-11-00 of the transient
X-ray pulsar GS 1843-02 performed at longitude
l=$\sim$31\deg and latitude 
b=$\sim$-0.5\deg in which the lowest intensity of the emission from this sky region was
recorded. The absence of $\sim$94s pulsations characteristic of this source and the characteristic shape
of the spectrum corresponding to the on state was
considered as an indicator that the pulsar was in the
off state. Thus, the flux recorded in this observation
corresponds to the flux of the ridge X-ray emission
proper.
  
  To additionally confirm the validity of the results
obtained, we analyzed observation no. 30416-01-01-02S of the NEW\_PULSAR\_NEAR\_SCUTUM region, in which GS 1843-02 in the off state fell within
the field of view of the spectrometer with a deviation of $\sim$45\arcmin from the optical axis and there were no
other sources in the field of view. As a result, we
obtained a spectrum similar to that measured when
analyzing observation no. 30141-05-11-00, but with
a lower statistical significance (due to a shorter exposure time). This confirms the correctness of the
chosen method for calculating the spectrum of the
ridge X-ray emission.
   
 It should be noted that in observation
no. 30141-05-11-00, apart from the pulsar
GS 1843-02, the pulsar PSR J1846-0258 was at
the edge of the field of view (the deviation from the
optical axis is $\sim$44\arcmin ) from which persistent radiation
was detected at a level of $\sim$2-2.3 mCrabm and which
is not transient. However, our estimates showed that
its contribution to the total flux recorded by PCA in
this observation did not exceed 10\%.
   
 We constructed the spectrum of the ridge X-ray
emission from all of these data with the intensity
recalculated for the latitude of the source under study
and then subtracted it from the combined spectrum
measured by PCA in the observations of Table 1.
 
   For the subsequent spectral analysis, we used
two spectra of XTE J1901+014 taken in the energy
range 3-20 keV by the method described above
from the PCA observational data of 1998 and 2002,
the combined hard spectrum taken with the ISGRI/INTEGRAL detector at energies above 20 keV
in 2003-2004, and the 0.6-2 keV PSPC-C/ROSAT
spectrum. For our timing analysis, we used
PCA/RXTE light curves with a time resolution of
16 s and 0.125 s to investigate the rapid variability
and ASM/RXTE data to investigate the long-term
variability.

\section*{LOCALIZATION AND THE OPTICAL COUNTERPART}

   XTE J1901+014 fell repeatedly within the
IBIS/INTEGRAL field of view in 2003-2004 during
deep observations of the Sagittarius Arm tangent.
This allowed the source to be detected at a high
confidence level ($>20\sigma$, Fig. 2a) and the accuracy of its localization to be improved considerably ($\sim$1.2\arcmin) compared to the ASM results. Note that
1RXH J190140.1+012630 also confidently falls into
the new error region of the source under study
(Fig. 2b), with the distance from the localization center of XTE J1901+014 to 1RXH J190140.1+012630
being no larger than 0.3\arcmin. The results obtained suggest
with a high probability that these two sources are
identical.
  
  The optical counterpart of the source under study
is difficult to find and identify, because it is close
to the Galactic plane. In particular, having analyzed the DSS (Digital Sky Survey) maps and
their own optical observations, Powell et al. (2002)
suggested that the blue star with the (J2000) coordinates R.A.= 19$^h$ 01$^m$ 39.$^s$90, DEC= +01\deg 26\arcmin 39\arcsec.2 (source \#1 in Fig. 3b) is the optical counterpart of the source. A further analysis of the 2MASS data
(http://irsa.ipac.caltech.edu/applications/2MASS)
revealed another source, 2MASS J19013983+
0126325 (source \#2), in the error region of
1RXH J190140.1+01263. This source is slightly
closer to the position of the X-ray source (cf. Figs. 3a,
3b, and 3c) and considerably brighter in the infrared
than source \#1 mentioned above (Fig. 3c). Note
that source \#2 is considerably fainter in the optical
bands and shows up only in the I band (Fig. 3b).
Comparative characteristics of the two presumed
counterpart stars are listed in Table 2.

\section*{LIGHT CURVES}

    First of all, note that no activity similar to the
1997 and 2002 events was found in the archival
ISGRI/INTEGRAL and PCA/RXTE data for
XTE J1901+014. Analysis of the ASM light curves
showed that no statistically significant flux was
detected from the source outside outbursts by this
instrument and there was no long-term variability.
  
  According to the ASM data, the flux from the
source during the outburst in June 1997 did not exceed the background level in the energy range 1.5-3 keV, was $\sim0.13$ Crab in the energy range 3-5 keV,
and reached $\sim0.7$ Crab in the energy range 5-12 keV.
This hard outburst was observed for 270 s, which
corresponds to three standard 90-s intervals of ASM
observations (Fig. 4a). The outburst in April 2002
was observed during two standard intervals of observations. The measured peak flux was higher than
that in the previous case, $\sim$1.1 and $\sim$1.2 Crab in
the energy ranges 3-5 keV and 5-12 keV, respectively.
The 1.5-3 keV flux from the source, $\sim$0.8 Crab, also
exceeded significantly the background level (Fig. 4b).
  
  As was noted above, no flux was recorded by
ASM from XTE J1901+014 in the quiescent state
because of its insufficient sensitivity. However, analysis of the source's observations with more sensitive instruments, the PCA spectrometer onboard the
RXTE observatory and the IBIS telescope onboard
the INTEGRAL observatory, has allowed us to detect
a weak persistent radiation from the source for the
first time. The intensity of this emission proved to be
the same both in the two series of RXTE observations of the source in 1998 and 2002 and during its
INTEGRAL hard X-ray observations in 2003-2004;
it was $\sim2.7$ mCrab in the energy ranges 3-20 keV and
17-100 keV, respectively.

    All of the PCA/RXTE observations show a certain temporal variability, a series of activity bursts
$\sim$40-60 s in duration with the peak flux ranging
from 7 to 9 mCrab (Fig. 5). No periodicity in the
occurrence of these bursts was found.

\section*{THE SPECTRUM}

\subsection*{The Quiescent State of the Source(Persistent Radiation)}

    Despite the relatively short PCA/RXTE exposure
time for XTE J1901+014 (see Table 1), we managed
to reconstruct high-quality spectra of persistent radiation from the object for both series of observations of
the source in 1998 and 2002. Comparison of the spectra obtained showed them to be virtually identical in
both shape and flux. This gave us grounds to analyze
the source's spectrum averaged over all observations.
The technique for properly reconstructing the spectra
of sources located near the Galactic plane is described
in detail in the "Observations" Section. The extent to
which the Galactic ridge X-ray emission affects the
actual spectrum of the source and the importance of
its allowance are demonstrated in Fig. 6. This figure
shows the combined PCA spectrum (Fig. 6a), the
spectrum of the Galactic ridge X-ray emission at the
latitude of the object under study (Fig. 6c), and the
true spectrum of XTE J1901+014 in the energy range
3-20 keV (Fig. 6b).
  
  We further analyze the spectrum obtained using
the XSPEC software package (http://heasarc.gsfc.
nasa.gov/docs/xanadu/xspec/index.html). In our
analysis, we took into account the 1\% systematic error related to the data reconstruction using software.
The spectrum was fitted by a power law,

$$ A(E)=K\times (E/1keV)^{-\Gamma}$$

where  $\Gamma$ is the photon index and $K$ is the normalization of photons/keV/cm$^{-2}$/s to 1 keV.
  
 Using the PSPC-C/ROSAT all-sky survey data,
we managed to obtain a statistically significant spectrum of the source in the energy range 0.6-2 keV
whose normalization closely coincides with that of
the PCA (3-20 keV) spectrum. However, to describe
the combined PSPC-PCA spectrum, apart from the
power law, the introduction of an interstellar absorption factor $M(E)=exp(-n_H\times\sigma(E))$, where $\sigma(E)$
is the absorption cross section (without Thomson
scattering), was required. According to our analysis,
this absorption toward the source in question corresponds to a hydrogen atom column density of $n_H=3.4\times10^{22}$ atoms cm$^{-2}$ . As a result, we obtained the
following best-fit parameters:$\Gamma = 2.26 \pm 0.03$, $K = 0.044 \pm 0.002$, $\chi^2 = 0.63$ (43 d.o.f.). The solid
line in Fig. 6b indicates the model described by these
parameters.
  
  The hard X-ray ($> 20$ keV) spectrum of the
source obtained from ISGRI/INTEGRAL data is
also described well by a power law with the same
slope. Given the constancy of the source's spectrum
in the energy ranges 0.6-2 keV and 3-20 keV as
well as the identical shape of the spectrum and fluxes
above and below 20 keV, it would be natural to
assume that the shape of the source's spectrum is
also constant over a wide energy range, 0.6-100 keV.
In this case, the broadband spectrum of the object
under study (Fig. 7) can be obtained by combining the
ROSAT, RXTE, and INTEGRAL data. For this purpose, in the XSPEC package, the ISGRI/INTEGRAL
spectrum was added to the PSPC-C/ROSAT and
PCA/RXTE spectra with a free normalization and
jointly fitted by a power law. This fit showed that
the spectra have not only identical photon indices,
but also identical normalizations. This confirms our
assumption about the constancy of the flux and
shape of the spectrum of persistent radiation from
XTE J1901+014 over a wide energy range.

    Analysis of the source's spectrum during its activity bursts in observations 30186-01-16-05 and
30186-01-21-01 revealed no changes in its shape;
only the flux varied.

\subsection*{OUTBURSTS (ASM data)}
    Based on ASM calibrations using PCA data,
Smith et al. (2002) showed that the slope for sources
with power-law spectra (and, in particular, for black
hole candidates) could be estimated using the ratio
of the count rates in two ASM hard energy channels
(3-5 keV and 5-12 keV):$$\Gamma_{ASM} = 1.499\times R + 0.698,$$             
where R is the ratio of the count rate in the 3-5 keV
channel to the count rate in the 5-12 keV channel.
 
   By recalculating the flux ratios for all 90-s intervals in which the source was detected by ASM,
we can roughly trace the evolution of its spectrum
during both outbursts. Whereas during the outburst
of July 1997 the source's spectrum was fairly hard
and virtually constant with ASM $\Gamma_{ASM}\sim 1.2$, in April 2002
the spectral slope changed from $\sim2.4\pm0.1$ in the first
ASM observing interval to $1.4\pm0.1$ in the second interval. Such a different spectral behavior of the source
may stem from the fact that the RXTE observatory
observed different phases of the outbursts in 1997 and
2002.

\section*{CONCLUSIONS}

    Our timing and spectral analyses of the intense
outbursts in 1997 and 2002 lead us to conclude that
these are most likely not the type I bursts associated with thermonuclear explosions on the surfaces
of neutron stars, because, in contrast to the latter,
either their hardness does not change with time or
they become even harder.
  
  A joint analysis of the INTEGRAL, RXTE, and
ROSAT data has allowed us to construct for the first
time a broadband (0.6-100 keV) energy spectrum of
the object under study, which is well fitted by a power
law with a photon index of $\sim$2.26. Neither cutoffs at
energies 20-30 keV, which are characteristic of X-ray pulsars (see, e.g., Filippova et al. 2005), nor, after
allowance for the contribution from the Galactic ridge
X-ray emission, any emission lines have been found
in the source's spectrum. Such a nonthermal spectrum without any evidence of a cutoff may indirectly
indicate that the compact object in the system under
consideration is a black hole.
  
 No long-term variability has been found in the
behavior of XTE J1901+014. However, short-term
aperiodic variability was detected in the energy range
3-20 keV as a series of activity bursts $40-60$ s in
duration whose peak flux exceeded the flux from the
source in the quiescent state by a factor of $1.5-3$.
The spectral shape of the source during these bursts
remains the same as that in the quiescent state.
  
 The intense outbursts     detected    from
XTE J1901+014 are similar in timing and spectral
characteristics to the well-known outbursts from
V4641 Sgr (Stubbings and Pearce 1999), Cygnus
X-1 (Golenetskii et al. 2003), and, to a lesser degree,
the outbursts from such fast X-ray transients as
SAX J1818.6-1703 (Grebenev and Sunyaev 2005).
The comparative parameters of the sources and their
outbursts are listed in Table 3.

   To summarize, we may assume that
XTE J1901+014 belongs to the class of fast X-ray transients with a black hole as the compact
object. However, further observations in various
wavelength ranges, primarily in the soft X-ray and
optical ranges, are required to determine the nature of
XTE J1901+014.

\section*{ACKNOWLEDGMENTS}

   We wish to thank E.M. Churazov, who developed the IBIS data analysis algorithms and provided
the software. We also wish to thank A.A. Vikhlinin,
M.G. Revnivtsev, and R.A. Krivonos for a discussion of the results obtained as well as S.S. Tsygankov and E.V. Filippova for help in the INTEGRAL
data reduction. We used data from the archive of
the Goddard Space Flight Center (NASA), the Integral Science Data Centre (Versois, Switzerland), and
the Russian Science Data Center for INTEGRAL
(Moscow, Russia). This work was supported by the
Russian Foundation for Basic Research (project nos.
05-02-17454 and 02-04-17276), the Presidium of
the Russian Academy of Sciences ("The Origin and
Evolution of Stars and Galaxies" Program), and the
Program of the Russian President for Support of
Scientific Schools (project no. NSh-1100.2006.2).
A.A. Lutovinov separately acknowledges the support
from the Russian Science Support Foundation.

\newpage

\section*{REFERENCES}


1. T. Belloni, G. Hasinger, and C. Izzo, Astron. Astrophys. {\bf 283}, 1037 (1994).

2. H. V. Bradt, R. E. Rothschild, J. H. Swank, et al.,
   Astron. Astrophys. {\bf 97}, 355 (1993).

3. E. V. Filippova, S. S. Tsygankov, A. A. Lutovinov,
   R.A.Sunyaev, Astron. Lett. {\bf 31}, 729 (2005).

4. S. Golenetskii, R. Aptekar, D. Frederiks, et al., Astrophys. J. {\bf 596}, 1113 (2003).

5. S. A. Grebenev and R. A. Sunyaev, Astron. Lett. {\bf 31},
   672 (2005).

6. K. Jahoda, C.B. Markwardt, Y. Radeva, et al., Astrophys. J. {\bf 163}, 401 (2006).

7. D. Karasev, A. Lutovinov, and S. Grebenev, in Proceedings
of the of 6th INTEGRAL Workshop "The Obscured Universe", 2006 in press (astro-ph/0611399).

8. F. Lebrun, J.P. Leray, P. Lavocat, et al., Astron. Astrophys. 441, 141 (2003).

9. A. A. Lutovinov, S. V. Molkov, and M. G. Revnivtsev, Astron. Lett. 29, 713 (2003).

10. C. Powell, A. Norton, C. Haswell, et al., Astron. Telegram 93 (2002).

11. R. Remillard and D. Smith, Astron. Telegram 88, 1
   (2002).

12. M. G. Revnivtsev, R. A. Sunyaev, D. A. Varshalovich,
   et al., Astron. Lett. 30, 382 (2004).

13. M. Revnivtsev, S. Sazonov, M. Gilfanov, et al., Astron. Astrophys. 452, 169 (2006).

14. V. Sguera, E. J. Barlow, A. I. Bird, et al., Astron.
   Astrophys. 444, 221 (2005).

15. D. Smith, W. Heindl, H. Swank, et al., Astrophys. J.
   569, 362 (2002).

16. R. Stubbings and A. Pearce, IAU Circ. No. 7253
   (1999).

17. R. Wijnands, Astron. Telegram 89 (2002).

18. C. Winkler, T. J.-L. Courvoisier, G. Di Cocco, et al.,
   Astron. Astrophys. 411, L1 (2003).


\newpage
\centering {
{\bf Table 1.  }{Instruments and observations}

\vspace{2mm}
\hspace{-15mm}\begin{tabular}{c|c|c|c}
\hline
\hline
Observatory/&Observation& Date of observation&Effective \\
 instrument&&&exposure time,ks\\

\hline
\hline

PSPC-C/ROSAT & RASS 3/17/51 & 17/09/1990 & $\sim$0.55 \\[1mm]
\hline
\hline
PCA/RXTE & 30186-01-21-01Z & 11/09/1998 & $\sim$0.120 \\[1mm]

PCA/RXTE & 30186-01-21-01A & 11/09/1998 & $\sim$0.12 \\[1mm]
PCA/RXTE & 30186-01-16-05S & 23/09/1998 & $\sim$1.2  \\[1mm]
PCA/RXTE & 70409-01-01-05Z & 21/04/2002 & $\sim$0.15  \\[1mm]
PCA/RXTE & 70409-01-01-05S & 21/04/2002 & $\sim$0.06  \\[1mm]
\hline
\hline
ISGRI/INTEGRAL & orbits 48-70 & March-May 2003 & $\sim$1400 \\[1mm]
ISGRI/INTEGRAL & orbits 121, 128, 131-135 & October-November 2003 & $\sim$219 \\[1mm]
ISGRI/INTEGRAL  & orbits 172-177, 185-188, 193 & March-April 2004 & $\sim$610 \\[1mm]
\end{tabular}

\newpage

{\centering
{\bf Table 2.  }{Possible counterparts and their magnitudes}

\vspace{0.5cm}
\begin{tabular}{c|c|c|c|c|c|c|c}
\hline
 star number & B & V & R & I & J & H & K \\ 
\hline
\hline
 1 & 21.16\a & 18.72\a & 19.65\a & 18.01\a/18.263\b & 16.439\b & - & - \\ 
\hline
 2 & -  & - & - & 18.1\b & 13.2\c & 11.3\c & 10.4\c \\ 
\hline
\end{tabular}

Note. The magnitudes in the table are given in accordance
with the following data: a --JKT (Jacobus Kaptyn Telescope,
Powell et al. 2002); b --the DENIS catalog (http://vizier.ustrasbg.fr/viz-bin/VizieR-3); c --the 2MASS catalog.
}

\newpage
\centering
\hspace{-20mm}{\bf Table 3. }{Comparative characteristics of XTE J1901+014 and known fast X-ray transients}

\vspace{2mm}
\hspace{-20mm}\begin{tabular}{c|c|c|c}
\hline
\hline
Comparative &XTE J1901+014&V4641 Sgr&SAX J1818.6-1703\\
characteristics &&& \\
\hline
\hline
Duration of outburst &$>2$ min $<3.15$ h &$\sim$16 h & $\sim$23 h \\[1mm]
&(06/04/2002)& (15/09/1999) &(09/09/2003) \\
\hline
Flux in the maximum &0.5-1.2 Crab&$\sim$12 Crab& $\sim$380 mCrab \\[1mm]
 &(1.5-12 keV)&(1.5-12 keV)&(18-45 keV) \\
\hline
Change in hardness  & H from $\sim$1 to $\sim$2.75  & H from $\sim$1.78 to& increased \\[1mm]
during out & at maximum & $\sim$2.82 subsequently & in the energy range \\
(H=ASM$_{5-12keV}/ASM_{3-5keV}$)& &from $\sim$2.82 to $\sim$1.21  &\\
\hline
Presence of other & 27/07/1997 & equally intense & 11/03/1998 \\[1mm]
similar outbursts &  & not observed & 09/10/2003 \\
&&&10/10/2003 \\
&&&(Sguera et al. 2005)\\
\hline
\hline
Presence of persistent &$\sim$2.7 mCrab & ? & not detected \\[1mm]
flux&(0.6-100 keV)&&at statistically \\
 &&&significant level\\
\hline
Variability of & short-term & intense &-  \\[1mm]
persistent flux &aperiodic&burst activity& \\
\hline
\hline
Type of object & - & binary system & -\\[1mm]
&&with black hole& \\
\hline
Companion Sp, m & ? & B9, $\sim$5-8 M$_{\odot}$ & B3 \\[1mm]
\hline
\hline
\end{tabular}

\begin{figure}[p]
\includegraphics[width=\textwidth,bb=35 403 570 710,clip]{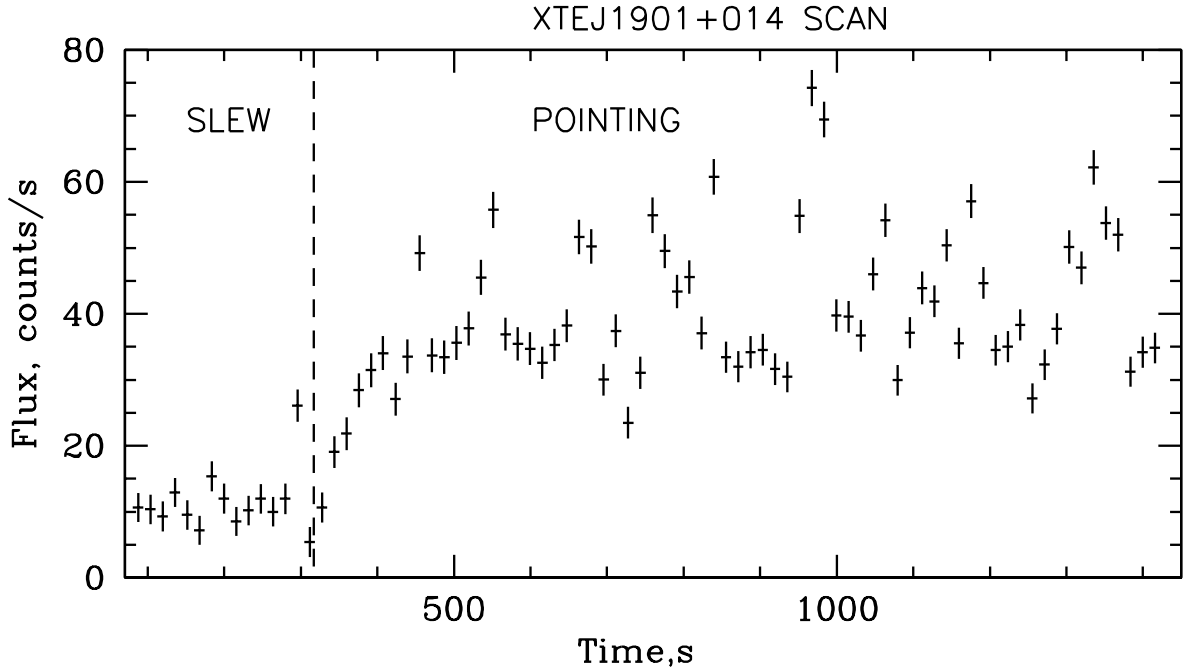}
\caption{ Light curve of XTE J1901+0.14 obtained from PCA/RXTE
observation 30186-01-16-05S, which reflects the variations in the
recorded 3╜20 keV flux related to the displacement of the PCA field of
view: slewing to the source and the source's observation proper at a
PCA/RXTE fixed field of view to the left and the right of the dashed
line, respectively.}

\end{figure}

\begin{figure}[p]
\includegraphics[width=1.01\textwidth,bb=13 12 1300 460,clip]{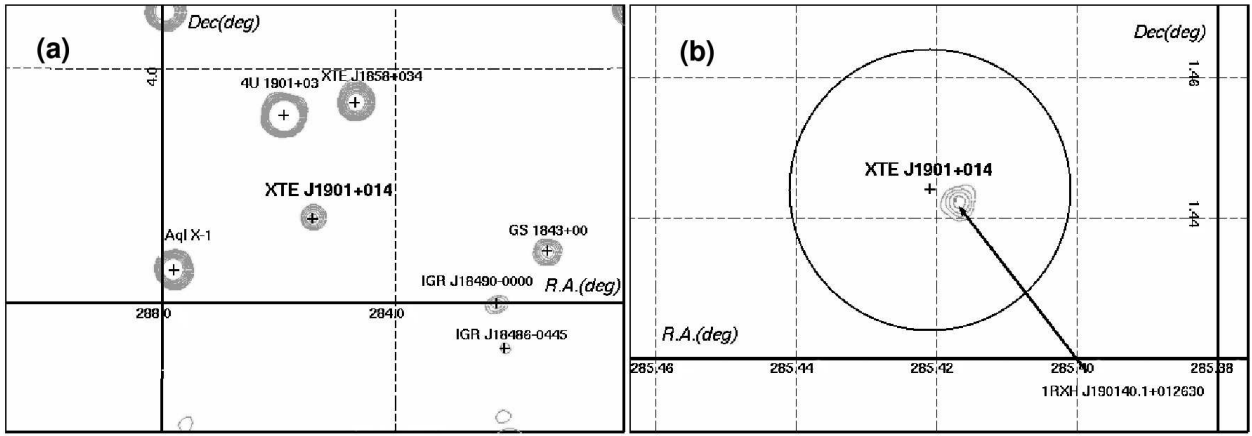}
\caption{Images of the sky region containing XTE J1901+014 obtained
from ISGRI/INTEGRAL data in the energy range 18╜100 keV (a) and from
HRI/ROSAT data in the energy range 0.1╜2.4 keV (b). The latter image
also shows the ISGRI/INTEGRAL localization center of the source and
the error circle of this instrument (with a radius of
$\sim$1.2\arcmin). The intensity contours indicate the position of
1RXH J190140.1+012630.}
\end{figure}

\newpage

\begin{figure*}
\vbox{
\centerline{\includegraphics[width=9cm]{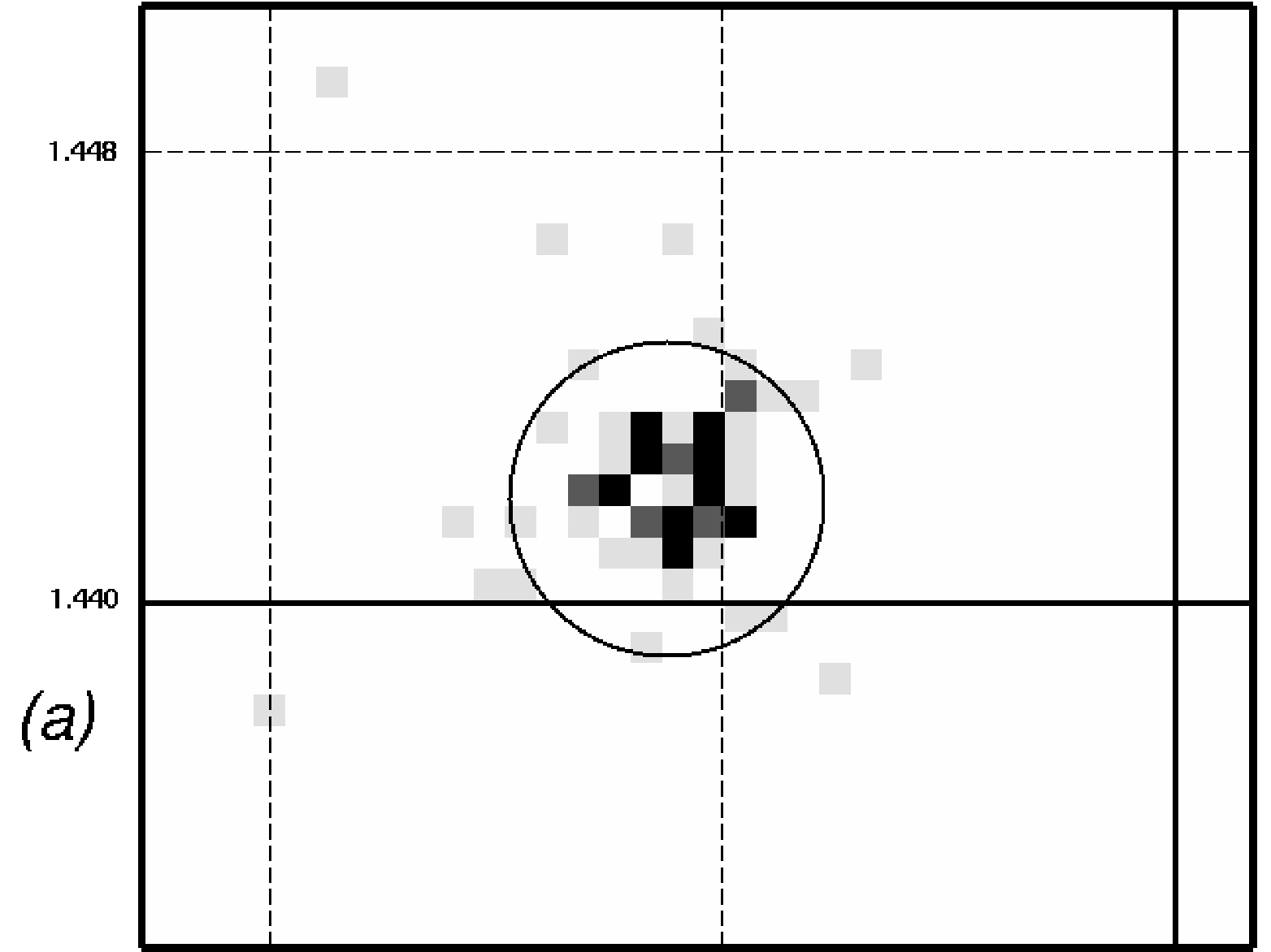}}
\centerline{\includegraphics[width=9cm]{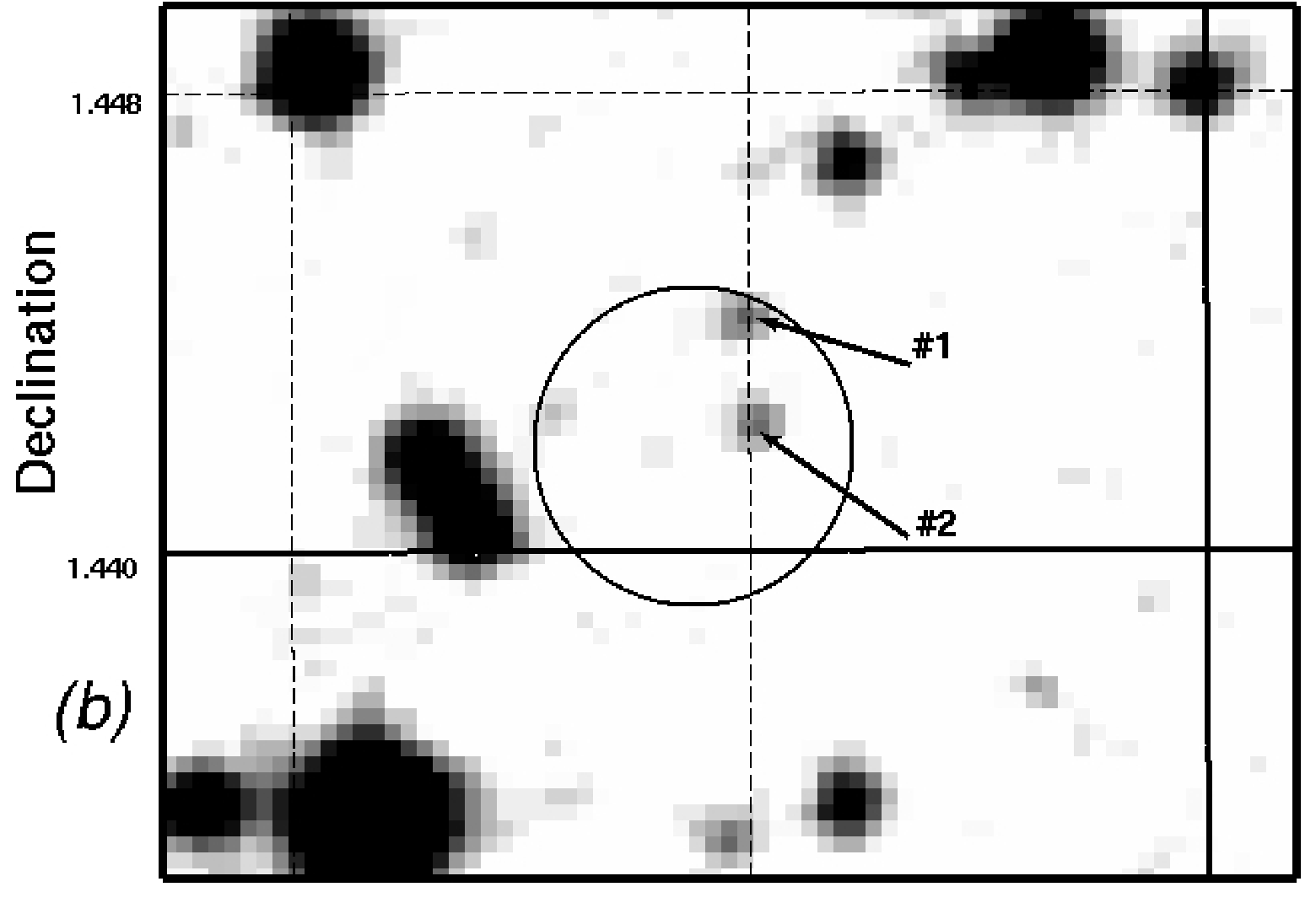}}
\centerline{\includegraphics[width=9.5cm]{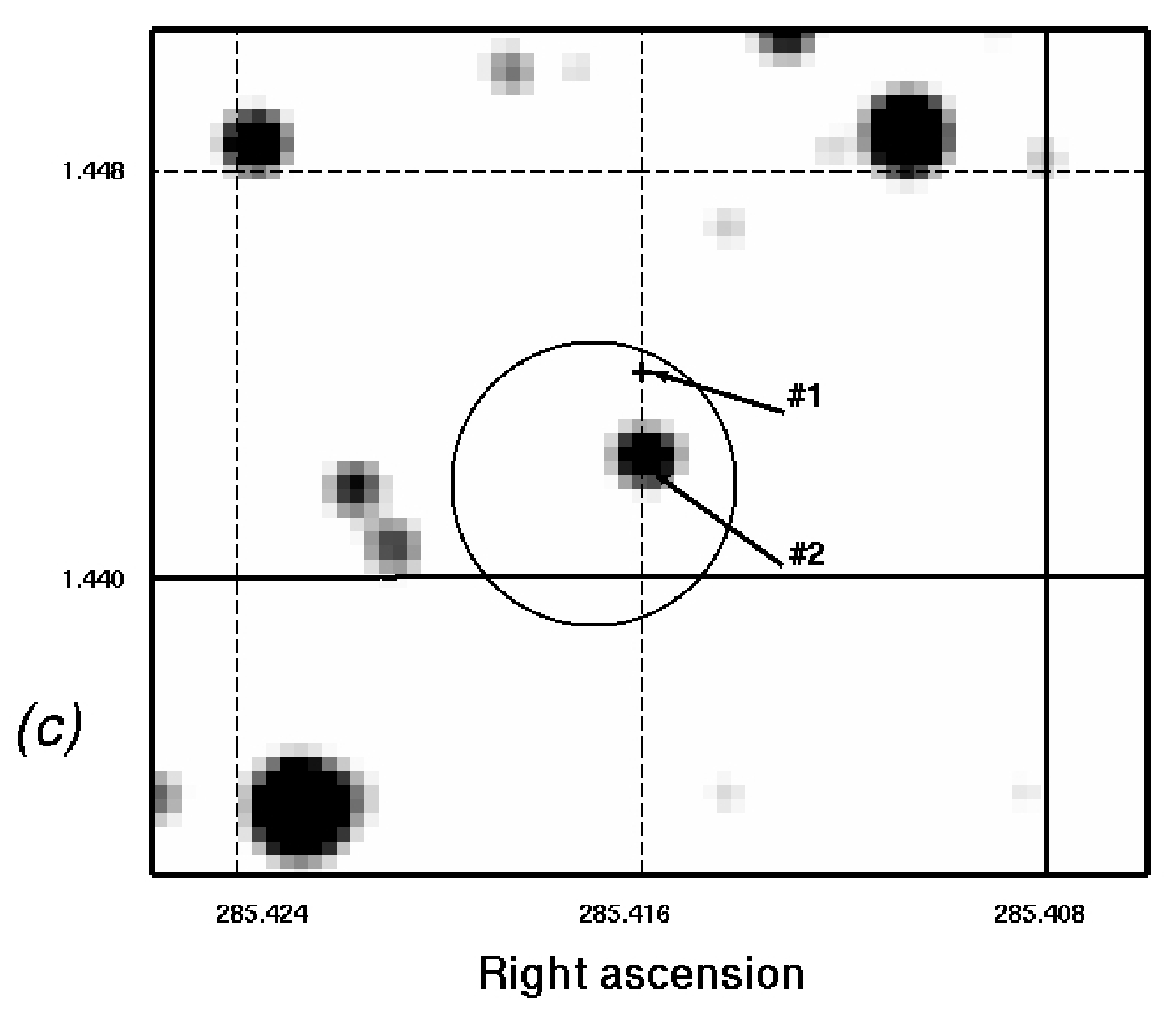}}
}
\caption{Images of the sky region containing
1RXH J190140.1+012630: (a) in the energy range
0.1╜2.4 keV based on HRI/ROSAT data, (b) in the I
band based on DSS data, and (c) in the J, H, and K
infrared bands based on 2MASS data. The localization
center and the HRI/ROSAT localization error radius
($\sim$10\arcsec) of 1RXH J190140.1+012630 are also marked in
the images. The numbers indicate the presumed optical
counterparts of 1RXH J190140.1+012630.}
\end{figure*}

\newpage
\begin{figure*}
\centerline{\includegraphics[ width=17.5cm ]{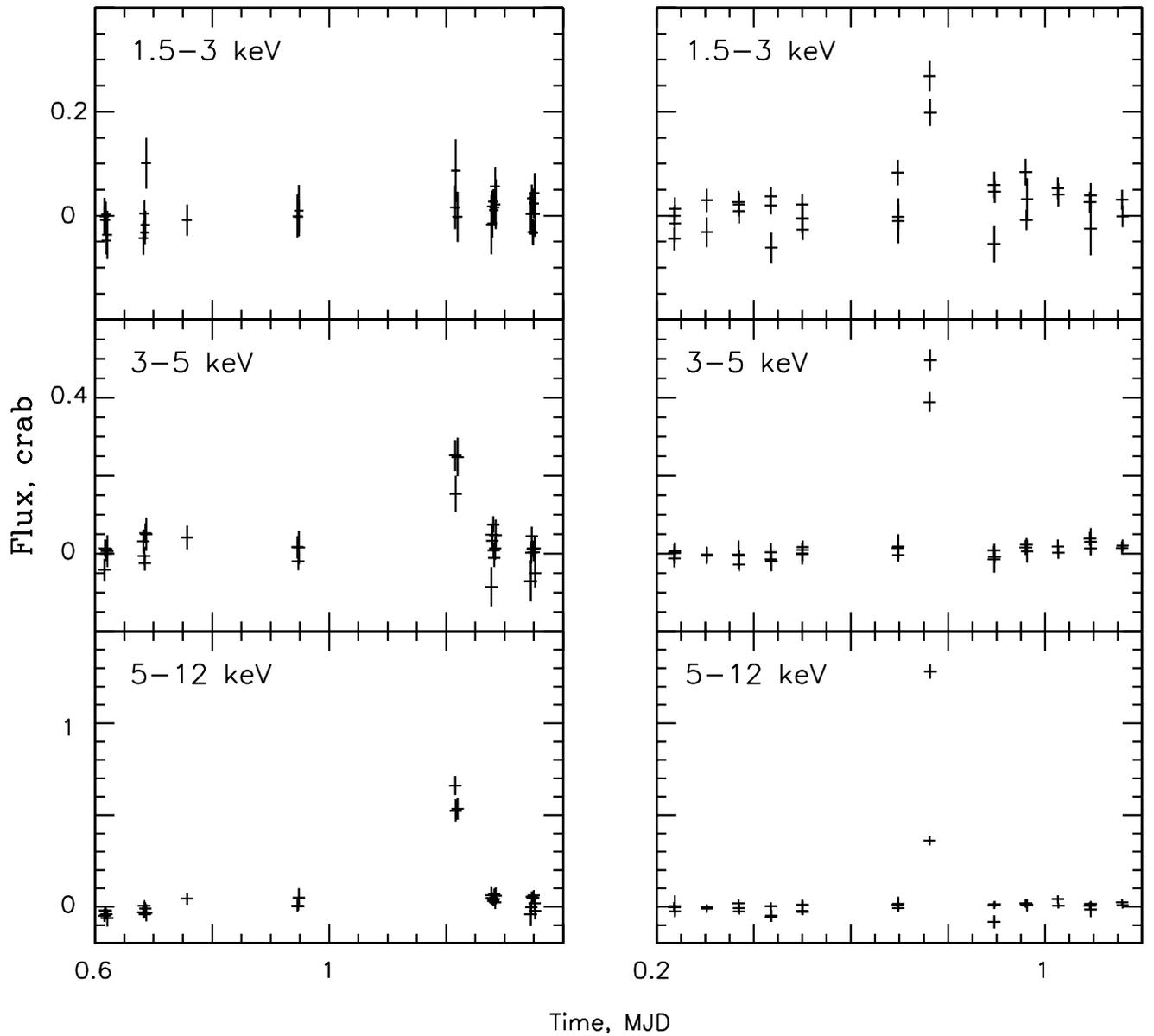}}
\caption{ASM/RXTE light curves of XTE J1901+014 during its outburst activity: (a) July 1997 (0 corresponds to MJD 50619,
UT 20.06.97 00:00:00); (b) April 2002 (0 corresponds to MJD 52370, UT 06.04.02 00:00:00). The duration of a single
observing time interval is 90 s.}

\end{figure*}

\newpage

\begin{figure*}
\includegraphics[width=17cm,bb=35 415 570 710,clip]{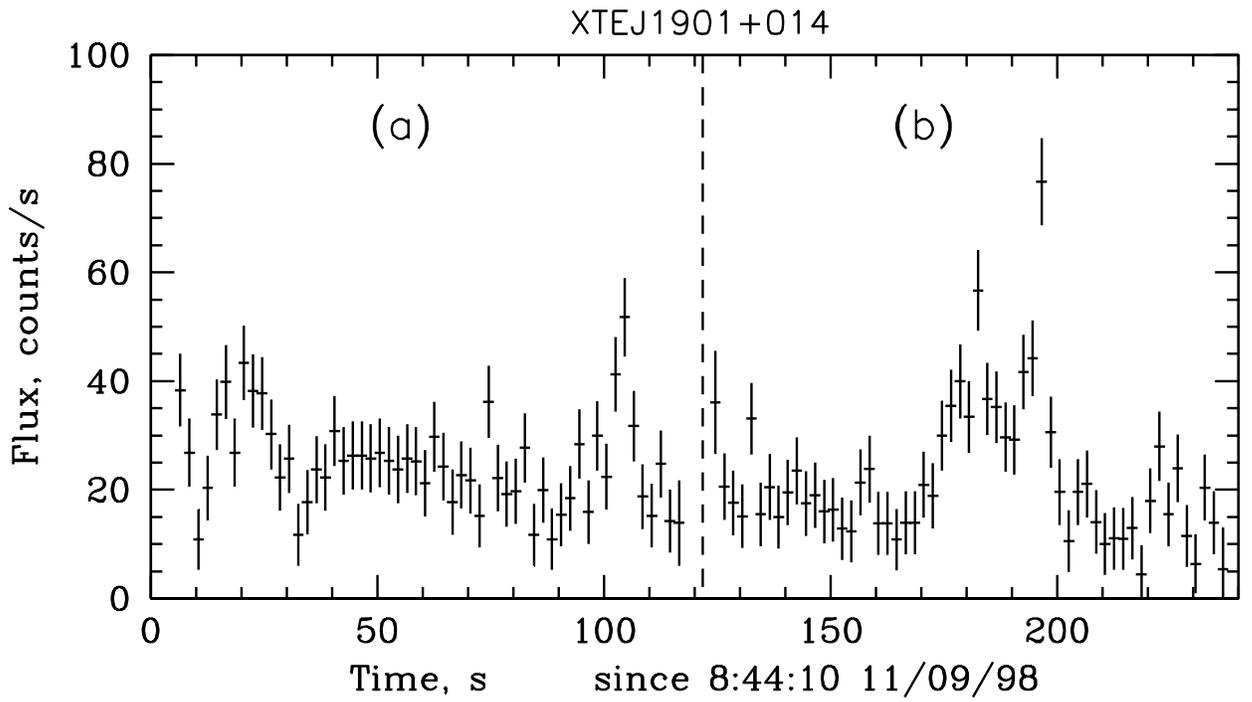}
\caption{ PCA/RXTE (3╜20 keV) light curves of
XTE J1901+014 for observations nos. 30186-01-21-01Z (a) and 30186-01-21-01A (b), which reflect the
source's aperiodic variability. The time resolution is 2 s.}
\end{figure*}
\newpage

\begin{figure*}
\includegraphics[width=17cm]{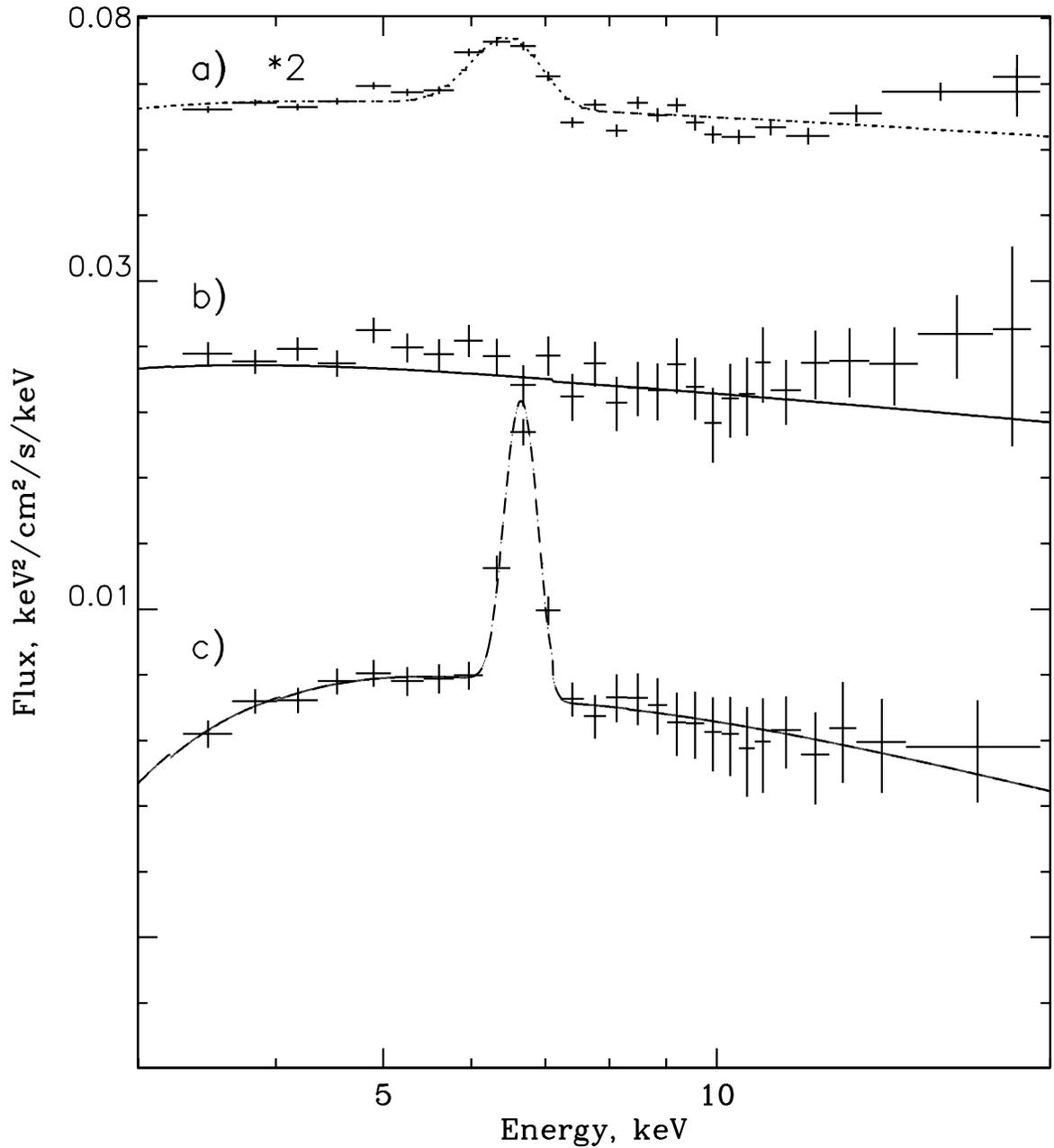}
\caption {(a) Energy spectrum of the sky region with a ra-
dius of 1 containing XTE J1901+014 constructed from
PCA/RXTE data (the intensity was doubled for clarity);
(b) the true energy spectrum of XTE J1901+014; (c) the
energy spectrum of the Galactic ridge X-ray emission.
The lines indicate the best fits to the spectra. }

\end{figure*}

\newpage

\begin{figure*}
  \includegraphics[width=17cm,bb=30 200 580 710,clip]{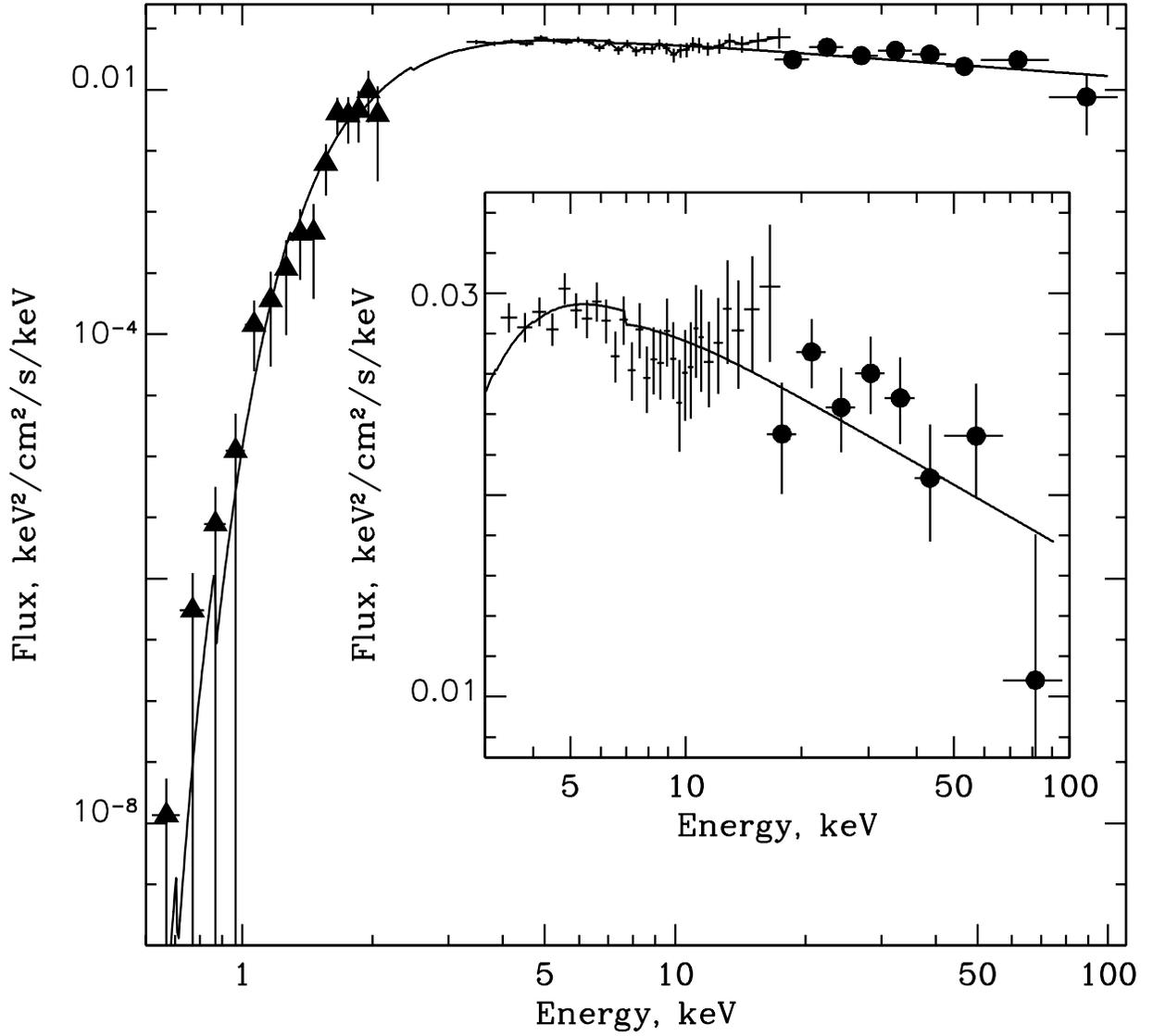}
\caption{Broadband (0.6-100 keV) energy spectrum of XTE J1901+014 based on PSPC-C/ROSAT (triangles), PCA/RXTE (crosses) and ISGRI/INTEGRAL (circles) data.}
\end{figure*}

\newpage

\end{document}